\documentclass[twocolumn,aps,prl,showpacs,amsmath]{revtex4}
\usepackage{graphicx}
\usepackage{bm,eucal}
\begin{document}
\title{The large-scale structure of passive scalar turbulence}
\author{Antonio Celani and Agnese Seminara}
\affiliation{CNRS, INLN, 1361 Route des Lucioles, 06560 Valbonne, 
France}
\date{\today}
\begin{abstract}
We investigate the large-scale statistics of a passive scalar
transported by a turbulent velocity field. At scales
larger than the characteristic lengthscale of scalar injection,
yet smaller than the correlation length of the velocity,
the advected field displays persistent long-range correlations 
due to the underlying turbulent velocity. These induce
significant deviations from equilibrium statistics for 
high-order scalar correlations, despite the absence of scalar flux. 
\end{abstract}
\maketitle 

Turbulent flows are systems far away from equilibrium,
characterized by a flux of energy 
through a range of scales. This cascade process is often accompanied
by strongly non-gaussian statistics and nontrivial scaling
properties \cite{F95}. Passive scalar turbulence is no
exception, in this respect. Its understanding has recently undergone 
remarkable progress, with results thoroughly reviewed in Ref.~\cite{FGV01}.
A passive scalar field $\theta({\bm x},t)$, 
like dilute dye concentration or temperature in
appropriate conditions, transported by an incompressible velocity
field ${\bm v}({\bm x},t)$, evolves according to
\begin{equation}
\partial_t \theta +{\bm v} \cdot {\bm \nabla} \theta= 
\kappa \Delta \theta + f\;,
\label{eq:1}
\end{equation}
where $f$ is a source of scalar fluctuations that acts at a
lengthscale $L$. The velocity field is assumed to be turbulent
and characterized by a self-similar statistics (e.g. 
$ (v(x+r)-v(x)) \sim r^{1/3}$ according to Kolmogorov's 1941 theory) 
in the range of scales delimited above by the velocity
correlation length $L_v$ and below by the viscous scale $\eta$.
Passive scalar fluctuations generated at the scale $L$ 
form increasingly finer structures due to velocity advection
and this process results in a net flux of scalar variance
to small scales, where it is eventually smeared out by molecular diffusivity
at a scale $r_d$. Here we will consider the case where these
scales are ordered as follows: $L_v \gg L \gg \eta,r_d$. In the range 
$L \gg r \gg r_d$ the average scalar flux is constant and 
equals the average input rate: 
this is the well studied inertial-convective range
where $\theta$ displays  non-gaussian statistics and
anomalous scaling \cite{SS00,W01}. Conversely, at scales larger than $L$
there is no scalar flux. Accordingly, one would expect 
Gaussian statistics and equipartition of scalar variance, i.e.
the hallmarks of statistical equilibrium. 
This expectation is correct at $r \gg L_v $, where 
the dynamics of the large-scale passive scalar
is ruled by an effective diffusion equation
resulting in a Gaussian statistics. 
However, in the intermediate range $L_v \gtrsim r \gtrsim L$,
deviations from 
``thermal equilibrium'' might arise as a consequence of
turbulent transport. Indeed, Falkovich and Fouxon \cite{FF04}  
have recently shown 
 ---  in the context of the Kraichnan model of passive scalar advection
where the velocity field is gaussian, self-similar 
and short-correlated in time \cite{K68}
--- that the scalar field shows a highly nontrivial behavior
at scales larger than the pumping correlation length $L$,
with significant differences from Gibbs statistical ensemble.
In this Letter we show that these results extend to 
passive scalar advection by a realistic flow, namely
two-dimensional Navier-Stokes turbulence in the 
inverse cascade range. This flow has been studied in great detail
both experimentally (in fast flowing soap films \cite{KG02} and 
in shallow layers of electromagnetically driven
electrolyte solutions \cite{T02}) and numerically \cite{SY93,BCV00}. 
The velocity is statistically homogeneous and isotropic, scale-invariant with
exponent $1/3$ (no intermittency corrections to Kolmogorov scaling) 
and with dynamical correlation times. 
This flow has also been utilized to investigate
passive scalar transport in the scalar flux range 
$L \gtrsim r \gtrsim r_d$ 
\cite{CLMV00,CV01} and multi-particle dispersion, 
an intimately related subject \cite{BC00,CP02}.
From this ensemble of studies it emerged that
the lessons drawn from the study of the Kraichnan model are indeed 
relevant for the qualitative understanding of passive scalar transport 
by realistic flows; and this will turn out to be true for the present case
as well.

\begin{figure}
\includegraphics[scale=0.7]{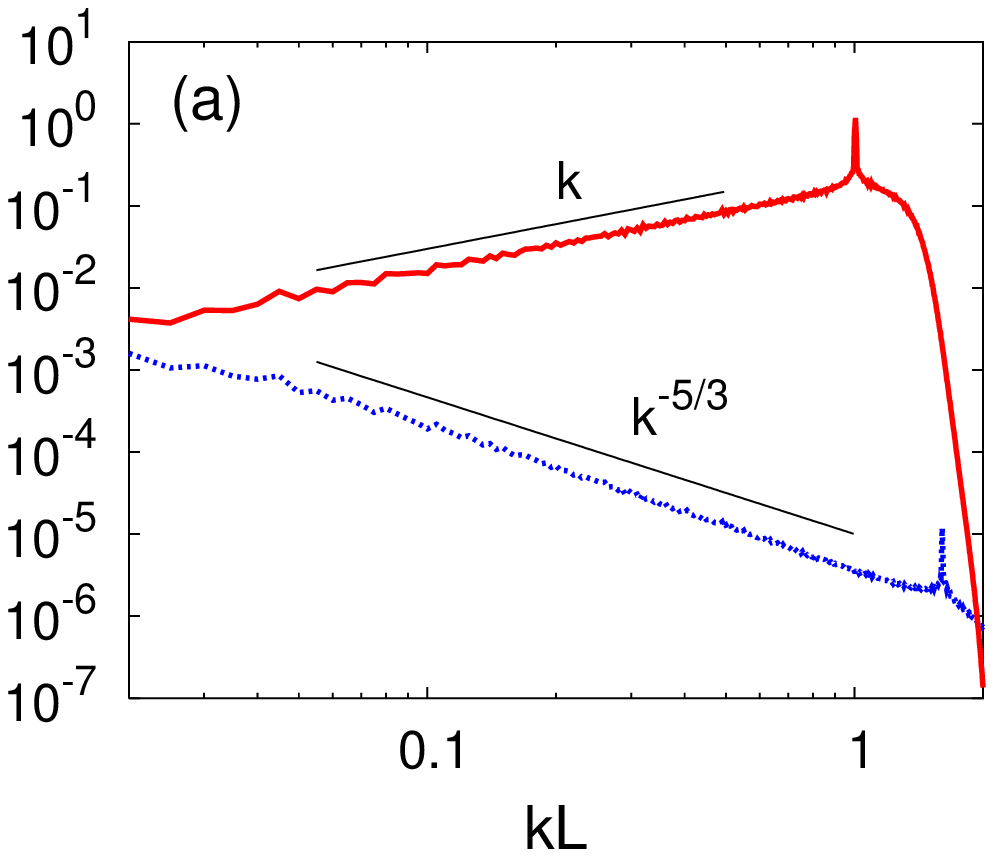}\\
 \includegraphics[scale=0.7]{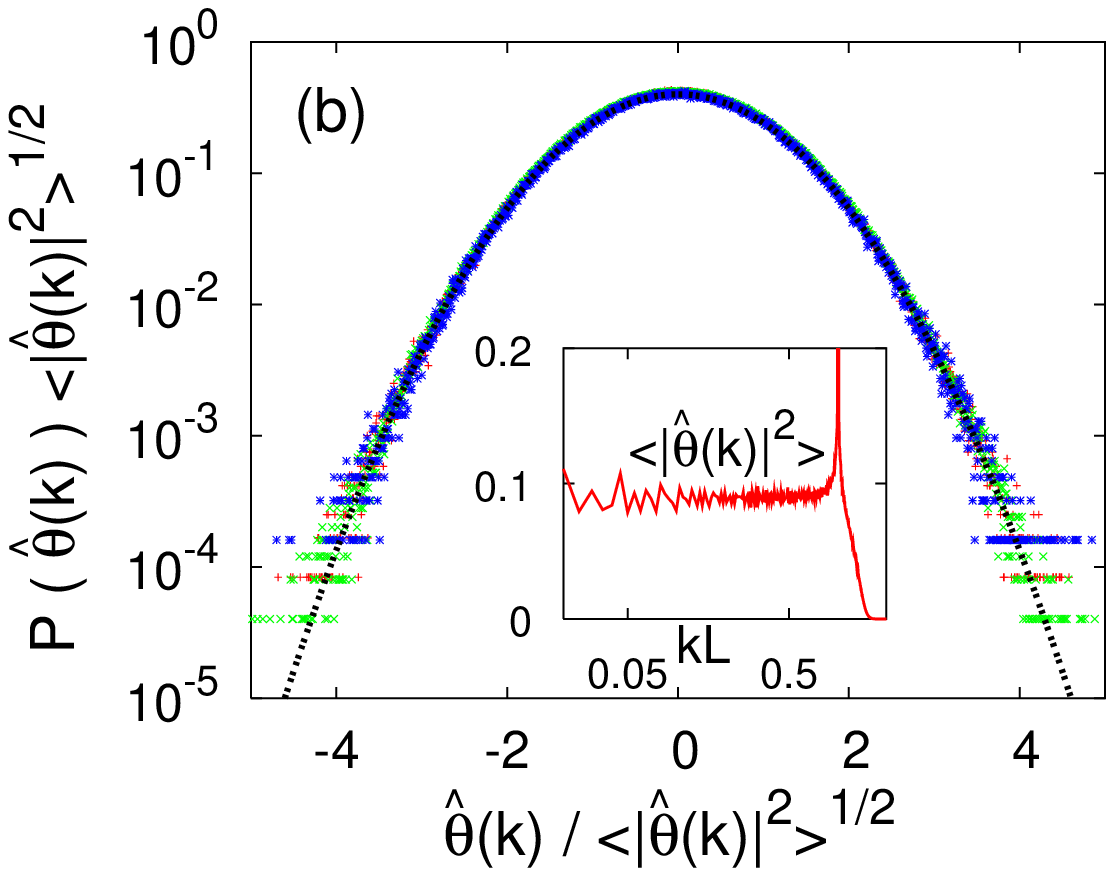}
\caption{(a) Passive scalar and velocity spectra. The data result from
the time integration of the two-dimensional Navier-Stokes equations
$\partial_t {\bm v} + {\bm v}\cdot {\bm \nabla} {\bm v} = -{\bm \nabla} p
+ \nu \Delta {\bm v} + {\bm F} -\alpha {\bm v}$
and eq.~\protect\eqref{eq:1} by a pseudospectral method on a 
1024$^2$ grid. The passive scalar is injected by a Gaussian, 
$\delta$-correlated in time, statistically homogeneous and isotropic
forcing restricted to a narrow band of wavenumbers.
The initial condition for the velocity field is a configuration
taken from a previous long-time integration 
and thus already at the statistically
stationary state. The passive scalar starts from a zero field 
configuration, and
after a transient of a few large-eddy turnover times
$\tau_v = L_v / v_{rms}$ where $L_v$ is the integral scale
of the velocity field, it reaches its own statistically steady state as well. 
Time averages are taken after this relaxation time has elapsed,
for a total duration of more than $10^4$ 
scalar correlation times $\tau_L \approx \tau_v (L/L_v)^{2/3}$. Here 
$L/L_v \approx 0.02$.
The velocity spectrum agrees with the Kolmogorov prediction $k^{-5/3}$
and the passive scalar one follows very closely the equipartition spectrum
in two-dimensions $E(k)\sim k$ (see also the inset of panel (b)).
(b) The marginal probability density function of a single Fourier amplitude
$\hat{\theta}({\bm k})$ is indistinguishable from a Gaussian (dotted curve)
for all wavenumbers in the range $L^{-1} \gtrsim k \gtrsim L_v^{-1}$. 
Here are shown
three wavenumbers with $kL=0.5,0.25,0.12$. In the inset is shown the 
spectral density $\langle|\hat\theta({\bm k})|^2\rangle$ that shows a neat plateau
at $kL \ll 1$ (notice the linear scale on the vertical axis).}
\label{fig:1} 
\end{figure}

Some basic, although incomplete, 
information about the scalar statistics in the supposed 
``thermal equilibrium range'' can be gained 
by studying the isotropic spectrum 
$E(k)=2 \pi k \langle |\hat\theta({\bm k})|^2\rangle$ for $ kL \gtrsim 1$. 
Let us recall that the equilibrium statistics for the scalar field 
would be described by the Gibbs functional 
${\cal P}[\hat\theta]=Z^{-1}\exp[-\beta\int|\hat\theta({\bm k})|^2 d{\bm k}]$, i.e. the Fourier modes should behave as independent Gaussian variables
with equal variance $1/(2\beta)$.   
As shown in Fig.~\ref{fig:1}, we observe $E(k) \sim k$, 
in agreement with equipartition
arguments; moreover, the statistics of single 
Fourier modes is indistinguishable from
Gaussian. 
However, we anticipate that from those findings alone one cannot
state conclusively that 
large-scale passive scalar is in a thermal equilibrium state,
given that they do not allow to rule out the possibility of
long-range correlations for higher order observables (e.g. four-point
scalar correlations).

\begin{figure}
\includegraphics[scale=0.7]{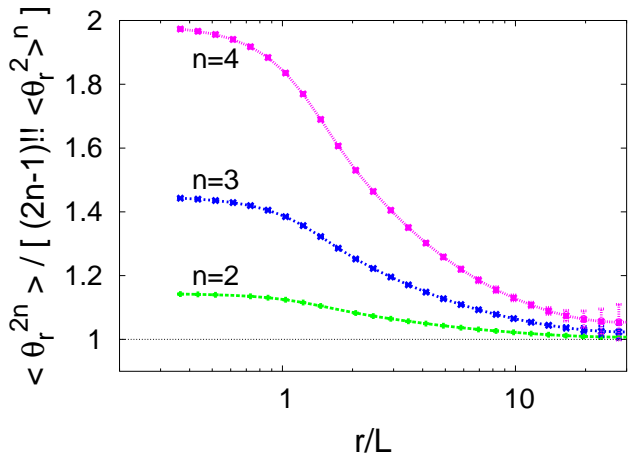}
\caption{The flatness and hyperflatness of the coarse-grained scalar field
as a function of $r/L$, normalized by their gaussian values $(2n-1)!!$.
For $r \to 0$ the curves tend to the flatness factors of the field 
$\theta$: the numerical values correspond to a supergaussian probability density function 
$\ln p(\theta) \sim -\theta^{1.6}$.}
\label{fig:2}
\end{figure}

A more refined description of the large-scale properties of passive scalar
can be obtained in terms of the coarse-grained field 
\begin{equation}
  \theta_r({\bm x},t) = \int G_r({\bm x}-{\bm y}) \theta({\bm y},t) \,d{\bm y}
\label{eq:2} 
\end{equation}
where $G_r$ acts as a low-pass filter in Fourier space (for instance,
the top-hat filter $G_r({\bm x} -{\bm y})= 1/(\pi r^2)$ if $|{\bm x} -{\bm y}|<r$ and zero otherwise; or the Gaussian filter 
$G_r({\bm x} -{\bm y})=(2 \pi r^2)\exp[-|{\bm x} -{\bm y}|^2/(2 r^2)]$).
For $r \to 0$ the filter reduces to a two-dimensional $\delta$-function
and therefore $\theta_r \to \theta$.
The statistics of $\theta$
is typically supergaussian \cite{JW91,JPG91}: 
its probability density function has exponential-like tails
even for a gaussian driving force $f$. Indeed, in the latter case
it can be shown that $\theta$ is the product of two
independent random variables $\theta \stackrel{\Delta}{=} \phi 
\sqrt{F_0 T}$ where $\phi$ is a gaussian variable 
of zero mean and unit variance, 
$F_0$  is the average injection rate of scalar fluctuations, and
$T$ is a positive-defined random variable, independent from $\phi$.
The variable $T$ is essentially
the time taken by a spherical blob of minute initial size to 
disperse across a length $L$ for a given flow configuration \cite{FMNV99,CV03}.
For example, assuming a poissonian distribution for the rare events when $T$ 
exceeds its average value yields exponential tails for
the pdf of $\theta$ \cite{SS94}.
The distribution of $\theta_r$ is supergaussian as well;
however, as $r$
increases above the forcing correlation length, the probability density
of $\theta_r$ tends to a gaussian distribution, as it is clearly seen
by the scale-dependence 
of the distribution flatness and hyperflatness (see Fig.~\ref{fig:2}). 

\begin{figure}
 \includegraphics[scale=0.7]{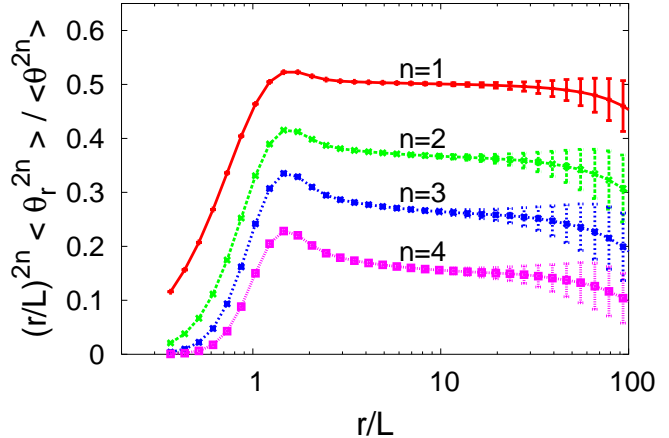}
\caption{Moments of the coarse-grained scalar field $\langle \theta_r^{2n} 
\rangle$ compensated by the thermal equilibrium expectation $(r/L)^{2n}$.
The errorbars are determined by dividing the sample in ten subsamples
and computing the dispersion around the mean.}
\label{fig:3}
\end{figure}

Within the framework of Gibbs statistical equilibrium, the
scalar field has vanishingly small correlations above the
scale $L$: therefore one could view $\theta_r$ as the sum of $N\approx(r/L)^2$
independent random variables (identically distributed as $\theta$) divided
by $N$. By central limit theorem arguments \cite{remark}, the moments of order 
$2n$ of the coarse-grained scalar
field (odd-order moments vanish by symmetry)
 should then scale as $N^{-n}$, giving $\langle (\theta_r)^{2n} 
\rangle \sim \langle \theta^2 \rangle^n (r/L)^{-2n}$. 
This is a very good estimate for $n=1$: indeed, as shown in  Fig.~\ref{fig:3},
the product $(r/L)^2 \langle \theta_r^2 \rangle$ has a very neat plateau.
This is consistent with the fast decay 
of the two-point scalar correlation $\langle \theta({\bm x},t)\theta({\bm x}+{\bm r},t) \rangle$ at $r \gtrsim L$. Indeed, in this case
the second-order moment $\langle \theta_r^2 \rangle = 
\int d{\bm y}_1 d{\bm y}_2 G_r({\bm y}_1-{\bm x}) G_r({\bm y}_2-{\bm x}) 
\langle \theta({\bm y}_1,t)\theta({\bm y}_2,t) \rangle $ 
is dominated by contributions with $|{\bm y}_1-{\bm y}_2| \lesssim L$ 
yielding  $\langle \theta_r^2 \rangle \sim \langle \theta^2 \rangle(r/L)^{-2}$.
Alternatively, by Fourier transforming the coarse-grained field one obtains
$\langle \theta_r^2 \rangle = \int  |\hat{G}_r({\bm k})|^2 
\langle |\hat\theta({\bm k},t)|^2 \rangle d{\bm k} 
\simeq \langle \theta^2 \rangle
(r/L)^{-2} $, since the transformed filter $\hat{G}_r({\bm k})$ 
is close to unity for $kr \ll 1$ and falls off
very rapidly for $kr \gtrsim 1$, and $|\hat\theta({\bm k},t)|^2 \simeq 
\langle \theta^2 \rangle/(\pi L^2)$. In summary, two-point statistics
appears to be consistent with Gibbs equilibrium ensemble. The situation
for multi-point correlations will turn out to be different.

A careful inspection of higher-order moments shows a 
less good agreement with central-limit theorem 
estimates (see Fig.~\ref{fig:3}): this 
points to the existence of subleading contributions to the moments
$\langle \theta_r^{2n} \rangle$ for $n>1$ arising from long-range
correlations of multiple scalar products. In order
to quantify more precisely the rate of convergence to gaussianity
and its relationship to long-range correlations,
it is useful to consider the cumulants of the random variable
$\theta_r$. According
to central limit theorem \cite{remark}, the cumulant of order $2n$ should
vanish with $N^{-2n+1}$ leading to an expected scaling 
$\langle\langle \theta_r^{2n} \rangle\rangle
\sim \langle\langle \theta^{2n} \rangle\rangle (r/L)^{-4n+2}$.
Let us reiterate that the former expression is expected to be valid
in absence of scalar correlations across lengthscales $r \gtrsim L$. 

\begin{figure}
\includegraphics[scale=0.7]{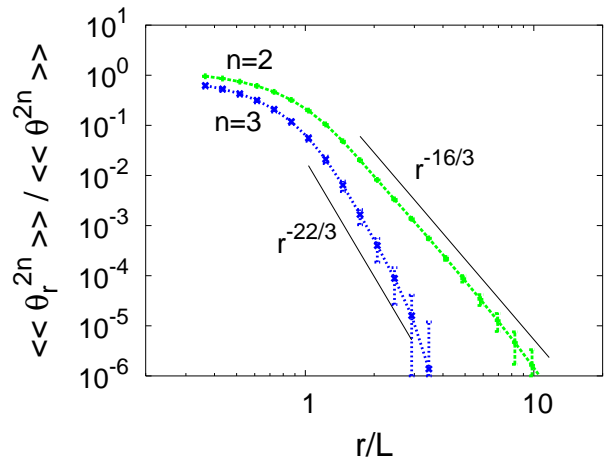}
\caption{Cumulants of order 4 and 6 for the coarse-grained scalar field. 
The best fits for the slopes give exponents
$5.32 \pm 0.05$ for $n=2$ and $7.5\pm0.5$ for $n=3$. 
The theoretical values $-16/3$ and $-22/3$ are shown for
comparison.}
\label{fig:4}
\end{figure}

For $n=1$ we have $\langle\langle \theta_r^{2} \rangle\rangle=
\langle \theta_r^{2} \rangle$ whose behavior has been already 
detailed above.
In Fig.~\ref{fig:4} we show the behavior of 
$ \langle\langle \theta_r^{4} \rangle\rangle = \langle \theta_r^4 \rangle
- 3 \langle \theta_r^2 \rangle^2$ and 
$ \langle\langle \theta_r^{6} \rangle\rangle = \langle \theta_r^6 \rangle
- 15 \langle \theta_r^2 \rangle \langle \theta_r^4 \rangle + 
30 \langle \theta_r^2 \rangle^3 $.
For the fourth-order cumulant, we observe a scaling law 
very close to the theoretical expectation  
$\langle\langle \theta_r^{4} 
\rangle\rangle \simeq \langle\langle  \theta^{4} \rangle\rangle 
 (r/L)^{-16/3}$ \cite{FF04}. 
This  has to be contrasted
with the scaling law 
$(r/L)^{-6}$ given by central limit arguments. 
The breakdown of central limit theorem 
is due to the existence of long-range dynamical
correlations in the range $r \gg L$.  These 
exclude the possibility of a true Gibbs 
statistical equilibrium at large scales.
The leading contribution to the fourth-order cumulant
$ \langle\langle \theta_r^{4} \rangle\rangle = 
\int d{\bm y}_1\, d{\bm y}_2\, d{\bm y}_3\, d{\bm y}_4\,G_r({\bm y}_1-{\bm x}) G_r({\bm y}_2-{\bm x}) G_r({\bm y}_3-{\bm x}) G_r({\bm y}_4-{\bm x})
\langle\langle \theta({\bm y}_1,t)\theta({\bm y}_2,t)  \theta({\bm y}_3,t)\theta({\bm y}_4,t)\rangle \rangle $ comes from configurations with 
the four points arranged in two pairs of close particles (e.g. 
$|{\bm y_1}-{\bm y_2}|\lesssim L$ and $|{\bm y_3}-{\bm y_4}|\lesssim L$)
separated by a distance $r$ (e.g. $|{\bm y_1}-{\bm y_3}| \simeq r$).
Otherwise stated, two-point correlators of the squared scalar field
$\langle\langle \theta^2({\bm x},t) \theta^2({\bm x}+{\bm r},t) \rangle\rangle$ must display a nontrivial scaling $(r/L)^{-4/3}$.
 We will get back to the issue
of the statistics of $\theta^2$
momentarily. The sixth-order cumulant $\langle\langle \theta_r^6 
\rangle\rangle$ is extremely
difficult to measure because of the strong cancellations between 
various terms. Upon collecting
the statistics over about ten thousand scalar correlation times, 
we can conclude that 
the results are consistent with the
power-law decay $\langle\langle \theta_r^{6} 
\rangle\rangle \simeq \langle\langle \theta^{6} 
\rangle\rangle (r/L)^{-22/3}$ suggested by the theory,
and arising from terms like $\langle\langle \theta_r^{4} 
\rangle\rangle \langle\langle \theta_r^{2} 
\rangle\rangle$ that appear in the expansion of the sixth-order 
cumulant \cite{FF04}.
The actual exponent for $\langle\langle \theta_r^{6} 
\rangle\rangle$ cannot be  determined with great precision, yet it lies 
within the range between $-7$ and $-8$, thus definitely different from 
the central-limit-theorem expectation $-10$.

\begin{figure}
 \includegraphics[scale=0.7]{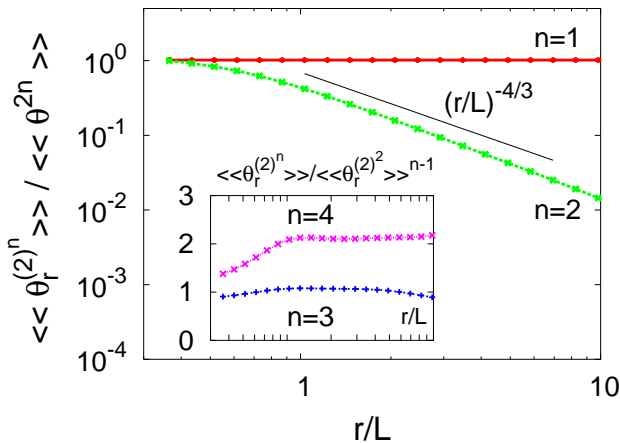}
\caption{Cumulants of the coarse-grained, squared scalar field $\theta^{(2)}_r$. The definitions for the low-order cumulants of a generic random variable $z$
 are: 
$\langle\langle z \rangle\rangle= \langle z \rangle$,
$\langle\langle z^2 \rangle\rangle= \langle z^2 \rangle - \langle z \rangle^2$,
$\langle\langle z^3 \rangle\rangle= \langle z^3 \rangle - 3\langle z\rangle
\langle z^2 \rangle + 2 \langle z\rangle^3$,
$\langle\langle z^4 \rangle\rangle= \langle z^4 \rangle-4\langle z\rangle
\langle z^3 \rangle-3\langle z^2 \rangle^2+12\langle z^2 \rangle \langle z \rangle^2-6\langle z\rangle^4$.}
\label{fig:5}
\end{figure}

Further insight on the deviations from
statistical equilibrium at large scales 
can be gained by studying the statistics of the coarse-grained
squared scalar field
\begin{equation}
  \theta^{(2)}_r({\bm x},t) = \int G_r({\bm x}-{\bm y}) \theta^2({\bm y},t) \,d{\bm y} \;.
\label{eq:3} 
\end{equation}
The cumulants of $\theta^{(2)}_r$ give useful information
about the presence of long-range correlations of the field
$\theta^2$. The first-order cumulant 
$\langle\langle \theta^{(2)}_r \rangle\rangle \equiv 
\langle \theta^{(2)}_r \rangle$
is trivially equal to $ \langle \theta^2 \rangle $. 
The second-order cumulant
$ \langle\langle {\theta^{(2)}_r}^2 \rangle\rangle =  \langle {\theta^{(2)}_r}^2 \rangle - \langle  \theta^{(2)}_r \rangle^2$ for a scalar field
in thermal equilibrium 
should decay rapidly to zero at large scales $r \gtrsim L$.
On the contrary, as shown in Fig.~\ref{fig:5}, 
we observe a slow power-law decay with an exponent
close to the theoretical expectation 
$ \langle \langle {\theta^{(2)}_r}^2 \rangle\rangle \simeq  \langle\langle 
\theta^4 \rangle \rangle (r/L)^{-4/3}$ \cite{FF04}.

\begin{figure}
\includegraphics[scale=0.4]{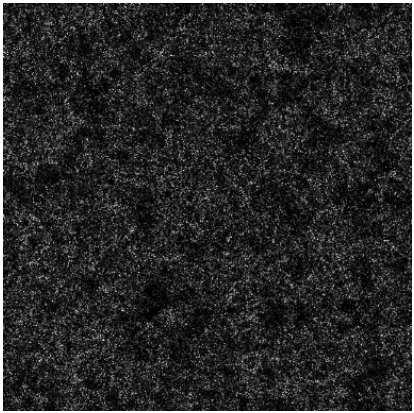}
\caption{A snapshot of the squared scalar field $\theta^2$. Remark
the inhomogeneous distribution of scalar intensity originating
from long-range correlations $\langle\langle{\theta^{(2)}_r}^2 \rangle\rangle
\sim r^{-4/3}$.  }\label{fig:6}
\end{figure}

Higher-order cumulants behave self-similarly as 
$ \langle\langle  {\theta^{(2)}_r}^n  \rangle\rangle \sim \langle\langle {\theta^{(2)}_r}^2  \rangle\rangle^{n-1}$. This result can be interpreted in terms
of the geometrical properties of the positive
measure defined by the squared scalar field: at scales $r \gtrsim L$ the field
$\theta^2$ appears as a purely fractal object with dimension $D_F \approx 2/3$ 
(see Fig.~\ref{fig:6}). We end up by briefly discussing the physical 
origin of long-range scalar correlations. For a gaussian forcing
we have $\theta^2({\bm x}_1,t)\theta^2({\bm x}_2,t) \stackrel{\Delta}{=}
F_0 \phi_1^2\phi_2^2 T_1 T_2$. At distances $|{\bm x}_1-{\bm x}_2|=r \gg L$
the two gaussian variables $\phi_1$ and $\phi_2$ are independent. However,
this is not the case for $T_1$ and $T_2$ because of the underlying
velocity field. Therefore, the long power-law tail
for  $\langle \langle {\theta^{(2)}_r}^2 \rangle\rangle$ arises
from events where $\langle T_1 T_2 \rangle \gg 
\langle T \rangle^2$. This amounts to say that
two blobs of initial size smaller than $L$, 
released at a distance $r \gg L$ in the same flow, 
do not spread considerably  by turbulent diffusion (i.e.
$T_{1,2} \gg \langle T \rangle$) with a probability $\sim (r/L)^{-4/3}$.

We acknowledge illuminating discussions with G. Falkovich and A. Fouxon.
This work has been supported by the EU under the contract
HPRN-CT-2002-00300.
Numerical simulations have been performed at CINECA (INFM parallel
computing initiative).


\begin{thebibliography}{99}
\bibitem{remark} 
The statistics of the random variable 
$z_N=(\sum_{i}^{N}w_i)/N$, where $w_i$ are independent and identically distributed random variables
with zero mean and finite variance $\sigma^2$,
is completely characterized by the generating function $G(s)=\langle \exp(sz_N)
\rangle$. By virtue of statistical independency and identity in distribution
of the $w$'s, we have $G(s)=\langle \prod_{i}^N \exp(sw_i/N) \rangle=\langle
\exp(sw/N)\rangle^N=g(s/N)^N$ where $g$ is the generating function for $w$.
For $N \gg 1$ we have $g(s/N) \simeq (1+s^2\sigma^2/(2N^2))^N \simeq \exp{s^2\sigma^2/(2N)}$ (a version of the central limit theorem) and therefore
$\langle z_N^{2n} \rangle \simeq (2n-1)!! \sigma^{2n} N^{-n}$.
The cumulants $\langle\langle z_N^{2n} \rangle\rangle$  are defined 
in terms of the Taylor series of $\ln G(s)$ around $s=0$: 
since $\ln G(s)=N \ln g(s/N)$ we have $\langle\langle z_N^{2n} \rangle\rangle
\sim \langle \langle w^{2n} \rangle \rangle N^{-2n+1}$.
\end{thebibliography}
\end{document}